\documentclass[11pt]{article}
\usepackage[margin=1.3in]{geometry}
\usepackage{amsfonts}
\usepackage{setspace}
\usepackage{hyperref}
\usepackage{enumitem}
\usepackage{verbatim}
\usepackage{natbib}
\usepackage{fancyhdr}

\fancypagestyle{plain}{%
  \fancyhf{}%
  \fancyhead[L]{Forthcoming in \emph{Erkenntnis}}
  \fancyhead[R]{Penultimate draft}
  \fancyfoot[C]{}%
}

\title{\vspace{-1cm}Big Bounce or Double Bang?\\\large{A Reply to Craig \& Sinclair on the Interpretation of Bounce Cosmologies}}
\date{\vspace{-5ex}}
\author{Daniel Linford}

\begin{document}

\maketitle

\section{Introduction}

We often hear that our universe began in a catastrophic event approximately fourteen billion years ago. Nonetheless, since the beginning of physical cosmology as a science in the first half of the twentieth century, physicists have explored ``bounce'' cosmologies (\cite{Kragh_2009, Kragh_2018}). According to the usual interpretation of bounce cosmologies, our universe originated when a pre-existing universe ``bounced'' through a highly compressed state; this could have happened in a variety of ways. The metaverse, of which our present universe is one proper part, might cycle through multiple generations of universes, each reaching a maximum size before contracting and eventually giving birth to a subsequent universe (as in, e.g., \cite{IjjasSteinhardt:2017, Ijjas_2018, Ijjas:2019pyf, Steinhardt:2002, steinhardt_2007}). Or there could have been a single previous universe that ``bounced'' through a maximally dense state to give birth to our universe, which will expand indefinitely into the future. (For reviews of models  in the former two families, see \cite{Lilley:2015ksa, Novello:2008ra, Battefeld:2015, Brandenberger2017}.) Alternatively, each universe might give birth to offspring universes through the highly compressed state found within black holes (\cite{Poplawski:2010, Poplawski:2016, Smolin:1992, Smolin:2006_CosmoSelection}).

William Lane Craig and James Sinclair disagree with the traditional interpretations of bounce cosmologies (\cite{CraigSinclair:2009, CraigSinclair:2012}). Craig and Sinclair defend the \emph{Kal$\overline{\textrm{a}}$m} Cosmological Argument for Theism: (1) every thing which begins to exist has a cause for its existence, (2) the universe began to exist, and, therefore, (3) the universe has a cause for its existence.\footnote{Though the \emph{Kal$\overline{\textrm{a}}$m} argument, as stated, is theologically neutral, proponents of the argument can either add supplementary arguments for the conclusion that God was the cause of the universe or use the \emph{Kal$\overline{\textrm{a}}$m} argument as one part of a cumulative case for God's existence (\cite{Draper:2010}).} To defend the second premise in light of bounce cosmological models, Craig and Sinclair re-interpret the interface between the two universes to represent the \emph{ex nihilo} birth of both universes -- a ``double big bang'' \cite[125-127]{CraigSinclair:2012}. While some philosophers of science have paid serious attention to the \emph{Kal$\overline{\textrm{a}}$m} argument's relationship to physics in decades past (e.g., \cite{Earman:1995, Smith:2000, Grunbaum:1989, Grunbaum:1991, Grunbaum:1996, Grunbaum:1998, Grunbaum:2000, Mortensen:2003, Pitts:2008}) and Craig and Sinclair's arguments crucially depend upon technical details from contemporary physics, close to nothing has been written in reply to the interpretations of the physical cosmological models appearing in \cite{CraigSinclair:2009} or \cite{CraigSinclair:2012}. Despite this, considerable attention continues to be given to Craig and Sinclair's articles in philosophy of religion. For example, according to a \emph{Google Scholar} search, \cite{CraigSinclair:2009} has been cited well over a hundred times. Craig and Sinclair's interpretations served as the basis for some of Craig's arguments in an important public debate with physicist Sean Carroll in 2014 r(printed as \cite{CarrollCraig:2016}); as of April 8, 2020, the recording of the Carroll/Craig debate has received more than 273,000 views on \emph{YouTube}. Therefore, a critique of Craig and Sinclair's treatment of cosmological models from the perspective of philosophy of science is sorely missing. Here, I take steps to rectify this situation. Setting aside theological considerations and questions about bounce cosmology's plausibility, I will show that bounce cosmologies have features which Craig and Sinclair's interpretation cannot plausibly explain. There are bounce cosmologies in which the features of one universe explain features of the other, which seems inconsistent with the interpretation that both universes were born simultaneously, and there are bounce cosmologies in which the thermodynamic arrow of time is continuous from one universe to the next.

\section{The beginning of the universe and the singularity theorems}

Before discussing bounce cosmologies, I need to place some conceptual machinery onto the table. First, in this section, I will briefly describe singularity theorems as the theorems apply to cosmology. Craig's defense of the \emph{Kal$\overline{\textrm{a}}$m} argument has often focused on one specific singularity theorem (the Borde-Guth-Vilenkin theorem, as described below), which Craig takes to provide evidence for the \emph{Kal$\overline{\textrm{a}}$m} argument's second premise, i.e., that the universe began to exist. There are cosmological models to which the singularity theorems do not apply, so that Craig and Sinclair's discussion of non-singular cosmologies, such as bounce cosmologies, revolves around how those cosmologies avoid the singularity theorems and whether non-singular cosmologies can avoid an absolute beginning. Second, in section \ref{interface_section}, I will briefly discuss the universe's entropy in order to describe Craig and Sinclair's interpretation of bounce cosmologies.

 The universe's expansion can be understood in terms of a characteristic length scale termed the \emph{scale factor} and denoted $a(t)$. For present purposes, it will suffice to say that the universe grows as $a(t)$ increases. Early in physical cosmology's history, physicists realized that models of an isotropic and homogeneous universe, with some assumptions about the matter-energy-momentum populating space-time, predict that $a(t)$ tends to zero at some finite time in the past. Imagine a time-line documenting the history of an isotropic and homogeneous universe.\footnote{In technical jargon, I am imagining a specific foliation of a Friedmann–Lema{\^i}tre–Robertson–Walker (FLRW) space-time into space-like surfaces. Undoubtedly, calling this a ``time-line'' is an oversimplification, but I ask for the reader's forgiveness for the sake of accessibility.} We'll need to pick a clock to label three-dimensional slices along our time-line. Let's choose a clock such that $a(t=0) = 0$. In that case, the three-dimensional slice labeled $t=0$ cannot be a part of the time-line because the space-time manifold is not mathematically well-defined when the scale factor is $0$; so, we need to remove $t=0$ from the time-line. Of course, different clocks will mark the slice that we remove with different labels, but every clock -- and so every observer -- will agree that there is a slice removed from the manifold. (To say this more precisely, every observer would agree that a space-like surface has been removed.) The three-dimensional slice that we've removed is an example of a \emph{curvature singularity}. Space-times from which points or slices have been removed, because the space-time manifold becomes undefined at the point or slice, are termed \emph{singular} space-times. With the three-dimensional slice removed, our time-line now consists of two half open segments. All of the slices ``after'' the removed slice can be placed into a sensible temporal order with respect to each other, but the slices ``prior'' to the removed slice no longer stand in any temporal relation to the slices ``after'' the removed slice. Therefore, we can disregard all of the times ``prior'' to the removed slice. Moreover, if we pick out any trajectory and follow that trajectory backwards in time, that trajectory must come to an end at the singularity because there is no space-time for the trajectory to traverse at, or before, the singularity. The singularity marks a boundary to space-time -- an absolute beginning. For Craig and Sinclair, a temporal boundary to space-time in the finite past is evidence for theism. As Craig describes singularity theorems, ``The standard Big Bang model [...] thus drops into the theologian's lap just that crucial premiss which, according to Aquinas, makes God's existence practically undeniable'' (\citet[238-9]{Craig:1992}). Elsewhere, Craig writes, ``What a literal application of the Big Bang model requires, therefore, is \emph{creatio ex nihilo}'' \cite[44]{Craig_Smith_1993A}.

As soon as the singular behavior of cosmological models had been discovered, physicists were suspicious. Perhaps cosmological models were singular as an artifact of assuming an unrealistic degree of homogeneity and isotropy, or perhaps the singularities in cosmological models were an indication that General Relativity would need to be replaced by a successor theory, as physicists already suspected on independent grounds. Physicists endeavored to provide theorems describing the conditions under which space-times are singular. Early theorems -- like those produced by Hawking and Penrose in the 1960s -- were able to show that curvature singularities were not the result of homogeneity or isotropy. (For a historical overview of singularity theorems up through the Hawking and Penrose theorems, see \cite{Earman:1999}.) In 2003, Arvind Borde, Alan Guth, and Alexander Vilenkin developed a new singularity theorem -- the BGV theorem -- with the advantage that the theorem no longer depended upon an explicit assumption about the universe's matter-energy-momentum contents (\cite{Borde:2003}). The BGV theorem applies to classical space-times generally, including space-times that are not solutions of the Einstein Field Equations.

To describe the BGV theorem, I first need to say what the Hubble parameter is and what geodesics are. The Hubble parameter can roughly be thought of as the universe's expansion rate and geodesics are the trajectories that particles traverse in space-time when no forces other than gravity act upon them. Time-like geodesics are the trajectories that particles with mass traverse while null geodesics are those traversed by massless particles. A geodesic that is neither time-like or null is termed `space-like'. A congruence of geodesics is a bundle of geodesics (analogous to a bundle of streamlines) filling a region of space-time and where no two of the geodesics cross. Borde, Guth, and Vilenkin develop a generalization of the Hubble parameter. We can think of the generalized Hubble parameter as the universe's expansion rate as measured by an observer traversing time-like or null geodesics. The BGV theorem is a result about congruences comprised by time-like and null geodesics. According to the BGV theorem, if the average of the generalized Hubble parameter along the geodesics comprising such a congruence is greater than $0$ -- that is, if a given observer traversing any of the geodesics in the congruence would observe the universe to be (on average) expanding along her geodesic -- then the congruence cannot be extended to past infinity. The termination of geodesics in the finite past within a given model is taken to be a strong indication that the model is singular, so that the BGV theorem suggests that all expanding space-times are singular. While Borde, Guth, and Vilenkin interpreted the singular behavior to indicate that our physical understanding is incomplete, Craig and Sinclair interpret the singular behavior as evidence for an absolute beginning. Nonetheless, as I will discuss in section \ref{DoubleBangSection}, a variety of non-singular cosmologies -- including bounce cosmologies -- have been proposed. Bounce cosmologies avoid an absolute beginning because instead of postulating that the generalized Hubble parameter is always greater than zero, bounce cosmologies postulate that space-time can be smoothly continued -- that is, without becoming singular -- from our expanding phase into a contracting phase. The interface at which the expanding and contracting phases smoothly join on to one another is termed the ``bounce''. 

To continue their defense of the \emph{Kal$\overline{\textrm{a}}$m} argument in the light of non-singular cosmologies, Craig and Sinclair have sought to provide a typology of cosmological models that ``evade'' the Hawking-Penrose or Borde-Guth-Vilenkin singularity theorems (\citeyear[143]{CraigSinclair:2009}; \citeyear[111]{CraigSinclair:2012}) and to show that either non-singular cosmological models suggest the universe did begin to exist or that non-singular cosmologies are implausible.\footnote{There are independent reasons to doubt that the BGV theorem tells us something significant about the origins of the totality of physical reality that I do not discuss here. Despite how the BGV theorem has sometimes been reported in the philosophy of religion literature, the BGV theorem is a result concerning the incompleteness of a congruence of time-like or null geodesics through a particular family of space-times as opposed to a more general result about the incompleteness of all of the time-like or null geodesics in a given space-time. Suppose that the average expansion rate along the time-like/null geodesics in the portion of space-time within our cosmological horizon is positive. If so, the BGV theorem tells us that those geodesics cannot be extended infinitely far into the past and remain within a classical space-time. Nonetheless, there could be time-like/null geodesics in regions beyond our cosmological horizon along which the average expansion rate is not positive. In that case, at least some time-like/null geodesics beyond our cosmological horizon could be extended infinitely far into the past. (That is, at least some time-like/null geodesics beyond our cosmological horizon could be complete, even if no geodesic within our cosmological horizon is complete, and the proper time measured along those geodesics could be infinite.) Guth has noted that eternally inflating models can lack a ``unique beginning'' and remain consistent with the theorem. Two time-like geodesics, along which the average expansion rate is positive and so cannot be extended infinitely far to the past, do not need to terminate at the same point or a common space-like surface. The theorem provides no upper bound to the lengths of \emph{all} of the time-like/null geodesics within the space-times to which the theorem applies \cite[6623]{Guth2007}. Andre Linde points out that, ``If this upper bound does not exist, then eternal inflation is eternal not only in the future but also in the past.'' As Linde continues, ``at present we do not have any reason to believe that there was a single beginning of the evolution of the whole universe at some moment $t = 0$, which was traditionally associated with the Big Bang'' \cite[17]{Linde:2007}. Moreover, even if all of the time-like/null geodesics within a given space-time were incomplete, the conclusion that there is an absolute beginning for all time-like/null geodesics does not follow from the statement that every time-like/null geodesic has a beginning. General Relativistic space-times can be sufficiently heterogenous as to preclude the possibility of defining an absolute beginning. Contrary to the \emph{Kal$\overline{\textrm{a}}$m} argument, the totality of physical reality would never have begun to exist. 

Craig and Sinclair briefly discuss this matter, (wrongly) interpreting it as an objection to the theorem (see footnote 41 in \cite[142]{CraigSinclair:2009}) and call the ``objection'' ``misconstrued''. Craig and Sinclair go on to assert that if the universe is eternal then if ``we look backward along the geodesic, it must extend to the infinite past if the universe is to be past eternal''. But this is false, as I've discussed; the point is that a space-time manifold can be geodesically incomplete -- in the sense proved by the BGV theorem -- without having an absolute beginning. Craig and Sinclair do admit that the BGV theorem is silent on what kind of singularity (or singularities) the metaverse contains.} In their typology, Craig and Sinclair discuss bounce cosmologies in which the entropic arrow of time reverses at the interface between universes; Craig returns to this point in his debate with Sean Carroll (\cite{CarrollCraig:2016}) and in discussion of Penrose's cosmological model (\cite[127]{CraigSinclair:2012}; \cite{CraigOnPenrose}). To complete my discussion of the requisite conceptual machinery for placing Craig and Sinclair's interpretation on the table, I turn to a discussion of the entropic arrow of time in the next section.

\section{\label{interface_section}The Interface and the Arrow of Time}

On the orthodox interpretation of bounce cosmologies, a preceding universe transformed into the highly compressed initial state of our universe. I will refer to the surface joining the two universes as the \emph{interface}. Craig and Sinclair disagree with the orthodox interpretation. To explicate Craig and Sinclair's re-interpretation of bounce cosmologies, I turn to a short digression on the status of the direction of time in fundamental physics. 

The fundamental laws of physics are \emph{time reversal invariant}. Consider a ball traveling at a fixed velocity in a vacuum. Now, suppose that the ball impacts and rebounds off of a wall. After the ball and the wall collide, in order to conserve momentum, the wall will begin traveling in the opposite direction. (We can suppose that the wall is resting on frictionless rollers and that only a negligible amount of energy was transferred into sound and heat on impact.) We can describe three events: (1) the ball is traveling at a fixed velocity while the wall remains at rest, (2) the ball and the wall collide, and (3) both the ball and the wall are traveling at fixed velocities in opposite directions. We can likewise define the time reverse: (1*) both the wall and the wall are traveling at fixed  velocities  towards one another, (2*) the ball and the wall collide, transferring all of  the wall's momentum to the ball, and (3*) the ball is traveling at a fixed velocity away from the wall while the wall remains at rest. Both sequences are equally allowed in Newtonian mechanics.\footnote{I've assumed that heat, friction, sound production, and other dissipative processes are  not part of Newtonian mechanics.} Newtonian mechanics is said to be time reversal invariant because, for all forward (reverse) sequences in Newtonian mechanics, the reverse (forward) sequence is nomologically permissible.  Quantum field theory -- and not Newtonian mechanics -- affords the best microphysical description of the actual universe. Nonetheless, quantum field theory is again time reversal invariant.\footnote{The sense in which quantum field theory is time reversal invariant is a subtle matter. Whenever we are provided with a microphysical description of the universe, in which we list sentences describing temporally sequential physical states, i.e., $S \equiv \{S_1, S_2, ..., S_N\}$, an operation can be constructed that is said to produce the time reversal of $S$, i.e., $S^* \equiv \{S_1^*, S_2^*, ..., S_N^*\}$. To say that the laws of physics are time reversal invariant is to say that both $S$ and $S^*$ are nomologically permissible sequences. Merely replacing $t$ with $-t$ in the equations of motion in fundamental physics does not suffice for time reversal. Instead, one must replace every charge  with the opposite charge, replace every system with its mirror image, and replace  every instance of $t$ with $-t$. That is, the fundamental laws respect the $CPT$ symmmetry and not the $T$ symmetry (\cite{Kobayashi_1973, Christenson_1964}). David Albert has argued that the fundamental laws have been known not to be literally reversible since the nineteenth century \cite[21]{albert_2000}; for a reply, see  \cite{Earman_2002}. In any case, the dynamical asymmetries that are known to appear in the fundamental laws do not explain the macroscopic asymmetries that appear in thermodynamics or in the special sciences. In those cases, the best explanation for the temporal asymmetry is offered by the reduction of time asymmetric phenomena to time symmetric phenomena in statistical mechanics.} 

Despite the time reversal invariance of the fundamental laws, macrophysical systems are obviously not time reversal invariant. We can fry an egg, but \emph{unfrying} an egg does not happen. How should the time asymmetry of macrophysical dynamics be explained? Phase space is the space of all possible microphysical states of a system. A given coordinate  in phase space represents a specific microphysical state. For any macrophysical observer, the exact microphysical state of a system is not accessible. For that reason, macrophysical descriptions carve up phase space into disjoint regions. If we conditionalize on the assumption that the universe began in a small region of phase space, then the most probable evolution of the universe's state is to another state that is a member of a larger phase space region. This affords a time asymmetric description from microphysical time symmetric dynamics.

To explain the macrophysically observable direction of time, a number of authors postulate that there was a time when the subspace of microphysical states consistent with the universe's macrophysical state occupied a vanishingly small phase space region (\cite{albert_2000, Albert:2015, Loewer_2007, Loewer:2012, Loewer:2020}). The size of a region to which a given microphysical state belongs is termed the \emph{entropy}.\footnote{This needs to be qualified. For the purposes of this paper, we can understand entropy as the hypervolume of a phase space region. The entropy is defined as the sum (or the integral) of $p_i \log(p_i)$, over the index $i$, where $p_i$ is the probability of occupying the $i^{th}$ microstate. The entropy will only be the hypervolume of the phase space region of interest on the assumption that the appropriate probability distribution is uniform over the phase space regions of interest, e.g., the Liouville measure.} The entropy has a macrophysical interpretation. Suppose that we would like a crowd to move a boulder that is too heavy for any individual to move. If we command all of the individuals in the crowd to charge at the boulder, but do not command them to coordinate their efforts, then, at best, the boulder will ``quiver'' when, by chance, more individuals charge the boulder on one side than on any other. The most effective way to have the crowd move the boulder involves the crowd coordinating their efforts, e.g., all of the individuals charging the boulder at a specific angle. There are a larger number of ways for the crowd to charge the boulder in a disorganized, uncoordinated fashion than for the crowd to coordinate their motions and so to charge the boulder in an organized fashion. 

Likewise, and for analogous reasons, there are fewer configurations of a gas that can do work in pushing a piston than there are configurations that cannot do work in pushing the piston. That is, at least from the perspective of nineteenth century thermodynamics, the entropy of a system is a measure of the system's ability to do macrophysical work. As the entropy of a system increases, the amount of energy available to do work decreases, and the system approaches an equilibrium state in which the system cannot do any work. For an engine to do work, there must be some reservoir of ``usable'' energy -- perhaps in the form of a temperature difference -- and when the reservoir has been depleted -- that is, when there is no longer a temperature difference -- the engine will have come to thermodynamic equilibrium and will be unable to do work. 

Contexts in which thermodynamics is combined with gravity and quantum mechanics, that is, contexts like the early universe, remain a matter of cutting edge and (often) speculative research. However, many physicists and philosophers of physics adopt the perspective that the early universe occupied a state far from equilibrium, characterized by low entropy, and the universe has been on a slow march towards equilibrium ever since. The asymmetry between the low entropy past and the high entropy future is thought to establish an entropy gradient along one temporal direction termed the \emph{entropic arrow of time}. Along the direction of the entropic arrow of time, the sequence of macrophysical events is ordered towards a state in the far future, when either no energy will be left for doing work and the universe will have reached equilibrium or the entropy will be ``re-set'', paving the way for a subsequent universe to begin in a low entropy state.

If this perspective is mistaken -- that is, if the early universe should not be described as occupying a low entropy state -- then the cosmological models that I offer later in this paper will, at the very least, need to be re-thought and much of what Craig and Sinclair conclude will have been shown to be unmotivated. Let's set that possibility aside and suppose that the early universe can be correctly described as occupying a low entropy state.

In some bounce cosmologies, the entropic arrow of time reverses at the interface between the two universes. On the orthodox interpretation, time has one direction through the bounce. That is, from the perspective of the expanding universe, the contracting universe is in the past. Craig and Sinclair disagree. Given the correlation between the direction of time and the entropic arrow of time, and that the entropic arrow points away from the interface in either direction, Craig and Sinclair argue that the interface should be understood as the birth of two universes (a ``double Big Bang''). As Craig and Sinclair describe, ``The boundary that formerly represented the `bounce' will now [be interpreted to] bisect two symmetric, expanding universes on either side'' \cite[122]{CraigSinclair:2012}. Elsewhere, Craig and Sinclair write that, ``The last gambit [in trying to avoid an absolute beginning], that of claiming that time reverses its arrow prior to the Big Bang, fails because the other side of the Big Bang is \emph{not} the past of our universe'' \cite[158]{CraigSinclair:2009}. As Craig and Sinclair conclude, ``Thus, the [universe on the other side of the interface] is not our past. This is just a case of a double Big Bang. Hence, the universe \emph{still} has an origin'' \cite[180-181]{CraigSinclair:2009}; also see \cite[125-127]{CraigSinclair:2012}.\footnote{Though much of the argumentation that Craig and Sinclair offer in their (\citeyear{CraigSinclair:2009, CraigSinclair:2012}) concerns the Aguirre-Gratton model (\citeyear{Aguirre:2002, Aguirre:2003}), Craig and Sinclair draw conclusions which Craig and Sinclair take to apply to \emph{any} cosmological model in which there is an interface at which the entropic arrow of time reverses direction, e.g., \cite[158]{CraigSinclair:2009}.}

One might worry that Craig's interpretation of the interface as a double Big Bang is inconsistent with claims Craig has made elsewhere about the irreducibility of the direction of time. The most robust defense of the view that the direction of time should be interpreted in terms of the entropic arrow is associated with a reductive program pursued by David Albert (\citeyear{albert_2000, Albert:2015}), Barry Loewer (\citeyear{Loewer_2007, Loewer:2012, Loewer:2020}), and David Papineau (\citeyear{papineau_2013}). \emph{Prima facie}, the Albert-Loewer-Papineau (ALP) reductive program is not consistent with Craig and Sinclair's theological project or Craig's metaphysics of time. For ALP, macrophysical temporally asymmetric phenomena should be given a reductive explanation. So, the temporal asymmetry of causal influence (e.g., effects cannot precede causes in time) can be explained, without remainder, in terms of phenomena that do not involve the time asymmetry of causal influence. This suggests that efficient causation may have a reductive explanation in terms of non-causal phenomena; if so, microphysical events need not have efficient causes.  As Alyssa Ney puts the point, ``from the point of view of microphysics, given an individual event, there is no objective distinction between which events make up that event’s past and which its future. Therefore, there is no microphysical distinction between which are its causes and which its effects. Thus, there are no facts about microphysical causation'' \cite[146]{Ney:2016}. If microphysical events do not require efficient causes, then not all events require efficient causes. So, contrary to Craig and Sinclair's theological aims, on ALP, the universe could have begun to exist uncaused.\footnote{Sean Carroll has offered a related argument. As Carroll points out, causal explanations typically depend upon the objects that stand in the explanation satisfying two conditions. First, that the objects obey the laws of physics and, second, a low entropy boundary condition in the past. The totality of physical reality is not an object \emph{within} physical reality and there is no known collection of physical laws that could apply to physical reality, as a whole, as opposed to applying to all  objects within physical reality. Moreover, there is no low entropy boundary condition beyond the totality of the physical world. Carroll concludes that we have no ``right to demand some kind of external cause'' for physical reality as a whole \cite[67-8]{CarrollCraig:2016}; also see \cite{Carroll:2005, Carroll:2012}.}

Of course, Craig and Sinclair need not sign on board to a reductive view of time or causation. In the metaphysics of time Craig favors, the direction of absolute time cannot be provided a reductive explanation in terms of the entropic arrow of time. Craig writes:

\begin{quote}
     From a theistic perspective [...] all such attempts [to reduce the direction of time] seem misconceived. For one can easily conceive of a possible world in which God creates a universe lacking any of the typical thermodynamic, cosmological or other arrows of time, and yet He experiences the successive states of the universe in accord with the lapse of His absolute time \cite[162]{craig_2001}.
\end{quote}

One can likewise imagine God experiencing the lapse of absolute time while the entropy of the universe decreases. In fact, one way that the universe \emph{could} lack the typical thermodynamic arrow of time would be if the entropic arrow of time and the direction of absolute time were not consistently aligned. None of this requires God's creation; one can replace God experiencing the lapse of absolute time with the metaphorical view-of-the-universe experiencing the lapse of absolute time. Craig has stated that if the entropic arrow and the direction of time do not align then this entails ``a non-reductionistic view of time [...] where the direction of entropy increase doesn't define the direction of time''. Craig objects that the misalignment between the entropic arrrow and the dirction of time ``is physically impossible'' because this would contradict the second law of thermodynamics \cite[78]{CarrollCraig:2016}. That is, that the alignment of the two is nomologically necessary. But Craig doesn't provide us with an account of why the alignment would be nomologically necessary. 

Importantly, the second law of thermodynamics is already known to be a statistical regularity that admits of exceptions. One advantage that the non-reductive view has is that time consistently flows in a fixed direction even when, e.g., through a statistical fluke, the entropic arrow reverses. If so, on the anti-reductionist view, the alignment between the entropic arrow and the direction of time is not nomologically, metaphysically, or logically necessary. In turn, if the direction of metaphysical time and the entropic arrow need not be aligned, then there is no reason to think the reversal of the entropic arrow of time, in those models where the entropic arrow of time reverses, suggests a double big bang. I think that this is a strong objection to Craig and Sinclair and I pursue a longer defense of this objection elsewhere (\cite{Linford:2020}). Nonetheless, I will set this worry aside so that I can pursue a different (and complementary) objection to Craig and Sinclair's interpretation of bounce cosmologies.

\section{A Double Big Bang?\label{DoubleBangSection}}

I can now turn to showing that Craig and Sinclair's interpretations of bounce cosmologies are implausible. If the interface were the \emph{ex nihilo} origin of two universes, then features of the universe on one side of the bounce, particularly those features that develop out of late time evolution, cannot provide an explanation for features of the universe on the other side of the bounce. But, as I will show, features of one universe do explain features of the other universe. Moreover, I will show that there are models in which the entropy is ``reset'' without reversing the entropic arrow.

\subsection{Anti-Inflationary Bounce Cosmologies}\label{AntiInflationSection}

The BGV theorem applies only to space-times with a positive average expansion rate. If a metaverse undergoes a contraction, the average expansion rate of the metaverse might not be positive. For one example, we can consider Anna Ijjas and Paul Steinhardt's pedagogical introduction to their favored family of bounce cosmologies (\citeyear{Ijjas_2018}). In the models that Ijjas and Steinhardt describe, a bounce between two universes is postulated in order to explain the features of our universe that inflationary cosmology had previously been meant to explain (e.g., the horizon, flatness, and smoothness problems); I will refer to these models as the anti-inflationary models. Consider the horizon problem. According to the horizon problem, in order for two regions, $R_1$ and $R_2$, of the Cosmic Microwave Background (CMB) to have reached a uniform temperature, signals traveling no faster than the speed of light would have had to have traveled from $R_1$ to $R_2$. But there are regions of the CMB that are further apart, on the Standard Big Bang (SBB) model, than signals could have traveled in the early universe.

Let's put some additional formal machinery on the table. For a given point $p$ in a relativistic space-time, the \emph{causal future} of $p$ is the set of points that a particle could reach from $p$ without exceeding the speed of light. The \emph{causal past} of $p$ is the set of points that could reach $p$ without having exceeded the speed  of light. And the \emph{absolute elsewhere} of $p$ are all of those points which cannot be reached from, and cannot reach, $p$ without exceeding the speed of light. Let's say that the \emph{patch} for $p$ at $T$ is the set of all the points at some time $T$ that are either in the causal future or causal past of $p$, where $p$ can be located at some time other than $T$. So, for example, there is a patch that consists of all of those points five minutes in my past from which particles can reach me now without exceeding the speed of light; included among those points are all of the points occupied by my computer five minutes ago, the entirety of the apartment that I am writing in five minutes ago, and so on.

We've said that, on the SBB model, when a present day observer, at time $t_1$, looks back to the early universe at $t_0 \ll t_1$, she can measure regions of the CMB between which signals could not have propagated without exceeding the speed of light. In other words, on the SBB model, when a present day observer, at time $t_1$, looks back to the early universe at $t_0 \ll t_1$, her patch at $t_0$ exceeds the horizon size at $t_0$. Nonetheless, she would observe her patch to be uniform in temperature, suggesting that the parts of the patch at $t_0$ must somehow have come into contact with one another. Inflation proposed a modification to the SBB model in which the early universe underwent a period of exponential expansion. If the universe underwent a period of exponential expansion, then the exponentially fast expansion of space could have pushed space-time regions that are initially in contact outside of one another's horizons. The anti-inflationary bounce cosmologies resolve the horizon problem in a different way. For anti-inflationary bounce cosmologies, the horizon of a previous universe was significantly larger than our patch in that universe. This would allow regions of the patch to come into thermal equilibrium \emph{before} our universe, so that the causally disconnected regions that produced the CMB would have uniform temperatures.

The anti-inflationary bounce cosmologies provide a natural explanation for the reduction in the entropy in the previous universe that allows a consistent arrow of time through the bounce. As Ijjas and Steinhardt describe, ``the patch corresponding to our observable universe today was only an infinitesimal fraction of the horizon size long before the bounce. That means only the limited entropy in the pre-bounce phase that is contained within [the patch] contributes to what is in the observable universe at the beginning of the expanding phase'' (\cite{Ijjas_2018}). Elsewhere, in describing ekpyrotic cosmological models\footnote{Craig and Sinclair have objected that the ekpyrotic model is geodesically incomplete and, therefore, not past eternal \cite[167-169]{CraigSinclair:2009}. However, Ijjas and Steinhardt have recently proposed a new version of the ekpyrotic model (\cite{IjjasSteinhardt:2017, Ijjas:2019pyf}). Steinhardt confirmed via correspondence that the new model can be made geodesically complete (per. corr. June 24, 2019).} -- one kind of anti-inflationary bounce cosmology -- Steinhardt and Turok write, ``Globally, the total entropy in the Universe grows from cycle to cycle [...]. However, the entropy density, which is all any real observer would actually see, has perfect cyclic behavior with entropy density being created at each bounce, and subsequently being diluted to negligible levels before the next bounce'' (\cite[1]{Steinhardt:2002}; also see \cite[192-193]{steinhardt_2007}). In stating that the ``global'' entropy grows from cycle to cycle, Steinhardt and Turok mean that the entropy generated within a given cosmological horizon during the previous cycle is not destroyed, but the entropy density is decreased exponentially, with an associated reduction of the degrees of freedom per horizon to nearly zero \cite[17]{Steinhardt:2002}. Thus, the entropy reversal through the bounce is not contrary to the entropic arrow of time; instead, the entropy simply left our causal horizon, rapidly becoming too distant for signals to successfully propagate to us.\footnote{\label{TolmanFootnote}In an argument that Steinhardt and Turok attribute to Richard Tolman the universe could not cycle through an eternity of contractions and expansions because entropy would build up in each cycle \cite[180-182]{steinhardt_2007}. As Helge Kragh has pointed out, ``Tolman did not actually conclude that there had been only a finite number of earlier cycles'' \cite[606]{Kragh_2009} and did not think thermodynamic considerations made a good case for the universe having begun at some finite time in the past. In fact, Tolman argued the universe might well extend infinitely far into the past and infinitely far into the future \cite[486]{Tolman_1934}. Nonetheless, as Steinhardt and Turok point out, the objection does not depend upon an increase in the total entropy; instead, the objection depends upon an increase in the entropy density in each cycle. For this reason, Tolman's argument is inapplicable to models in which the entropy becomes dilute (\citeyear[192-193]{steinhardt_2007}) or becomes hidden behind a horizon.}

Anti-inflationary bounce cosmologies are inconsistent with Craig and Sinclair's interpretation for two reasons. First, anti-inflationary bounce cosmologies resolve the horizon, smoothness, and flatness problems by invoking features of the late time evolution of another universe, which would be difficult to explain if one universe did not precede the other in time. As Steinhardt and Turok describe, ``The events that occured before the big bang shaped the large scale structure of the universe observed today, and the events that are occuring today will determine the structure of the universe in the cycle to come'' \cite[xiv]{steinhardt_2007}. Second, the anti-inflationary bounce cosmologies explain the low entropy of the early universe in a way that consistently maintains the entropic arrow of time through the bounce. As anti-inflationary bounce cosmologies have (arguably) become the most popular bounce cosmologies, and Craig and Sinclair's interpretation is inconsistent with the anti-inflationary bounce cosmologies, we may already have reason to reject their interpretation altogether. But let's push forward.

\subsection{Bouncing Through Black Holes}

The BGV theorem applies only to classical space-times. Models that modify the Einstein Field Equations to produce a non-classical space-time can produce a non-singular ``bounce'' (\cite{Poplawski:2010, Poplawski:2016, Starobinsky:1980, Sotiriou:2010, Corda:2011, Edholm:2018, Lilley:2015ksa, Kehagias:2014, IjjasSteinhardt:2017}). In this section, I consider two models -- Lee Smolin's evolving universe scenario (\citeyear{Smolin:2006_CosmoSelection, Smolin:1992}) and Nikodem Poplawski's model (\citeyear{Poplawski:2010, Poplawski:2016}) -- in which Einstein's gravity is modified in ways that allow the interior of a black hole to ``bounce'' and produce a baby universe.\footnote{Quentin Smith has offered a different, but related, cosmology in which universes are born from black holes \cite{Smith:1990, Smith:2000_BH}. I do not consider Smith's model in this paper because his model is not a bounce cosmology and the parent/child universes do not bear a temporal relationship to each other.} In Smolin and Poplawski's models, the thermodynamic arrow of time is continuous along geodesics that pass through the interface even though the entropy is ``reset'' at the interface.

Smolin's evolving universe hypothesis was developed to explain the so-called \emph{anthropic coincidences}. That is, that the free parameters appearing in our best theories of fundamental physics (e.g., the cosmological constant, the coupling constants, and so on) are consistent with the existence of life -- or large scale structures generally -- only if the parameters assume values from a narrow range compared to the range of values that the parameters could have had. That is, if the parameters are understood to be selected from a prior distribution uniform over all possible values of the parameters, then life is improbable. This problem can be resolved if one can provide a non ad hoc and plausible hypothesis according to which the probability distribution over the space of possible parameter values is not uniform. In other words, while the uniform distribution would poorly predict the existence of life, the existence of life is a prediction of the new distribution induced by the theory. Smolin's model is an attempt to provide one such explanation \cite[173-174]{Smolin:1992}.

If the density of matter or energy within some volume is sufficiently large and the matter-energy density outside the region sufficiently low, the result is a black hole. Like the models of the Big Bang discussed previously, General Relativistic models entail that black holes contain a curvature singularity. And just as with the Big Bang, physicists suspect that black hole curvature singularities will be replaced in a quantum mechanical description. Smolin's evolving universe hypothesis consists of two postulates:

\begin{enumerate}
    \item The curvature singularities General Relativity predicts to reside inside black holes will be replaced by the beginning of a child universe within a complete quantum gravity theory.
    \item In the creation of a child universe, the values of the free parameters in fundamental physical theories will slightly change (\cite[175]{Smolin:1992};  \citeyear[6]{Smolin:2006_CosmoSelection}).
\end{enumerate}

From these two postulates, given that large universes produce large numbers of black holes, large universes will have many more offspring than small universes. But the universes cannot be too large, or otherwise matter can never clump together to form black holes. The size of a given universe is determined by the rate at which the universe expands. Thus, the two postulates entail that the metaverse will come to be dominated by universes selected from a fixed range of expansion rates. In turn, the expansion rate is determined by the cosmological constant. So, a restriction on the range of expansion rates entails a restriction on the range of cosmological constants. The consequence will be that universes with values of the cosmological constant no larger than some maximum value would come to dominate the metaverse. Smolin has argued that his hypothesis affords an explanation of most of the other anthropic coincidences. For example, the hypothesis explains the difference in mass between the proton and neutron and provides an explanation for the gauge hierarchy problem (\cite{Smolin:1992}). Moreover, Smolin's hypothesis makes a falsifiable prediction that could be used, in principle, to rule out the hypothesis. Because universes with values of the free parameters that maximize the number of black holes would dominate the metaverse, we should predict that variations of the parameters characterizing our universe would result in universes with fewer black holes \cite[176]{Smolin:1992}.
 
Smolin's model requires that at least two features of the offspring universes be explained in terms of features of the parent universes. First, the model's ability to deliver on the desideratum that the hypothesis provide a non ad hoc reason for thinking that the distribution on the range of possible parameter values is not uniform. The evolving universe hypothesis satisfies this desideratum by entailing that the majority of the distribution's mass is located within the range conducive to black hole production.  If we begin with some population of $n$ universes, such that $n \geq 1$, selected from a distribution uniform over the space of possible free parameter values -- or, indeed, a variety of other distributions -- this distribution will generically evolve to a situation in which most of the probability mass \emph{is} located within the range of life-conducive values.\footnote{Two caveats are in order. First, the reader should not take the language of ``beginning'' too seriously. Smolin's model is consistent with a metaverse with an indefinitely long history and does not require that the metaverse ever \emph{began} to exist. Second, if we begin the model with $n=1$ universes and the cosmological constant selected for the initial universe is too large, the one universe could accelerate apart without producing any black holes. Therefore, in order for Smolin's model to work, either the iniital universe must be sufficiently improbable so to produce some population of black holes or else we should consider a situation in which we begin with multiple universes. Presumably, the most sensible possibility would be a metaverse that is eternal into the past so that there has always been some network of universes connected by black holes.}

Second, Smolin's model involves dynamics that maximize the number of black holes a given universe produces. Therefore, Smolin's model predicts that, if we vary the measured values of the free parameters characterizing our current universe, we should find that variations would result in hypothetical universes that would produce fewer black holes \cite[176]{Smolin:1992}. This prediction is explained by a selection history of prior universes. Consequently, if the interface between parent and child universes is not understood as the birth of a child universe out of a parent universe, then Smolin's model cannot do the explanatory work the model sets out to do.

Given that at least two features of the offspring universes must be explained in terms of features of the parent universes, one should not interpret the interfaces between universes that appear in Smolin's model as the birth of two universes. Instead, one should favor Smolin's interpretation, in which parent universes give rise to offspring universes.

I now turn to discussing Poplawski's cosmology. Poplawski's cosmology is produced within the Einstein-Cartan framework. Einstein-Cartan is a modification to the Einstein Field Equations that results from coupling spin -- an intrinsic property of some fundamental particles -- to torsion -- a geometric property of space-time. Coupling spin-to-torsion prevents the formation of singularities in black holes. Instead of forming a singularity, the black hole creates a child universe. The entropic arrow of time is continuous from the parent universe, through the black hole, and into the child universe (\cite{Poplawski:2010, Poplawski:2016}).\footnote{Poplawski’s model has an advantage over some other bounce cosmologies, because Poplawski's model avoids one of the criticisms Craig and Sinclair leverage against bounce cosmologies. As Craig and Sinclair argue, models in which a previous universe collapses to some minimum size before expanding into our universe needs to be carefully fine-tuned from eternity past in order to successful collapse to the minimum size (\citet[111-2]{CraigSinclair:2012}). But, like Smolin's model, Poplawski's model involves the creation of offspring universes from black holes. For this reason, neither Smolin's nor Poplawski’s models require such fine-tuning. Between the two, Poplawski's model is more convincing because, unlike Smolin, Poplawski provides a mathematical model and a physical mechanism for the dynamical evolution of black holes within one universe into subsequent offspring universes.} One might worry that bounce cosmologies in which the arrow of time is continuous through the interface do not avoid a beginning because the entropy could not have been increasing from eternity past. Consider, for example, counting backwards from ten. If one counts one number per second, then, after ten seconds, one must reach zero. So, if the entropy decreases into the past, shouldn't we hit some absolute zero on the entropy scale at some finite time in the past?

Here, the answer is no, and for two reasons Poplawski offers in his (\citeyear{Poplawski:2010}). First, while the field equations for Einstein-Cartan gravity are time symmetric, the boundary conditions of black holes are not time symmetric. That is, objects can travel through the black hole's event horizon but cannot travel back out. For this reason, the boundary condition for the child universe would be temporally asymmetric. Second, while observers outside the black hole will observe the horizon of the black hole maximizing the entropy, the entropy will not have been maximized for observers inside the black hole -- that is, in the offspring universe -- for whom the entropy can be increased still further. An entropy gradient requires only that the  entropy on the interface be smaller than the present entropy, but not that the entropy has never been lower than  the entropy on the interface. Indeed, there exist monotonically increasing functions $f(t)$ such that, for any time $T_0$, there will exist some $T_{-1} < T_1$, such that $f(T_{-1}) \leq f(T_0)$. For this reason, an entropy gradient can be established without postulating a beginning.\footnote{For a reply to a related argument originally offered by Tolman, see footnote \ref{TolmanFootnote}. As in Steinhardt and Turok's model, in the offspring universe, the total entropy of the parent universe has become hidden behind a horizon and is not accessible to the offspring universe.}

\subsection{Conformal Cyclic Cosmology}

Roger Penrose has proposed a different modification to General Relativity that, again, avoids the BGV theorem by proposing a non-classical space-time.\footnote{According to Craig and Sinclair, Penrose's model is consistent with the BGV theorem because the average expansion rate of a cyclic universe is zero \cite[111]{CraigSinclair:2012}. This is incorrect. In the CCC, the expansion or contraction of the universe is not a well-defined notion for every period of the universe's evolution. But, during those periods in which expansion/contraction are well-defined, the universe only expands and never contracts. The CCC is not singular because the CCC utilizes a space-time to which the BGV theorem cannot be applied.} Though the CCC is not typically considered a bounce cosmology, I will include the CCC in this paper for two reasons. First, Craig and Sinclair offer an interpretation of the CCC that parallels their interpretation of bounce cosmologies. Importantly, Craig and Sinclair argue that the interface between universes that appears in the CCC should be interpreted as the birth of two universes because the interface is an entropy minimum (\cite[127]{CraigSinclair:2012}; also see Craig's blogpost (\citeyear{CraigOnPenrose})). Second, several features of the CCC bear a significant resemblance to features of models traditionally considered bounce cosmologies (e.g., cylic generations of universes, an entropy minimum on the interface between universes, one universe that results in a highly compressed state in order to produce a subsequent universe).

Penrose postulated the CCC to explain the low entropy of the early universe (\cite[144]{penrose:2012}). As Penrose notes, the most probable way for a universe to evolve \emph{into} a cosmologically-relevant curvature singularity results in a highly disordered state because the evolution generically involves an amplification of any anisotropies or inhomogeneities. The anisotropries and inhomogeneities are amplified into black holes and then the black holes successively fuse. Penrose argues that this complex singularity structure attributes high entropy to the gravitational degrees of freedom. General Relativity is a time reversal invariant theory, so that the most probably way to evolve into a cosmological singularity should be the time reverse of the the most probable way to have evolved from a cosmological singularity. Consequently, the most probable way to evolve from a cosmological singularity would again involve a complex singularity structure which attributes high entropy to the gravitational degrees of freedom (\cite[124-125]{penrose:2012}).\footnote{\label{singularity_footnote}This can be put more carefully. As I noted earlier in the paper, curvature singularities are not points that General Relativity includes in the space-time manifold. Therefore, one should not say that, in General Relativistic models, the universe began with a low entropy singularity. However, one can accurately say that, when the universe is reversed in time in General Relativistic models, space-time tends towards a low entropy singularity. If one chooses any arbitrarily small value $\varepsilon > 0$ then there will exist some time $t$ such that the scale factor $a(t) < \varepsilon$. The low entropy singularity corresponds to the limit in which $\varepsilon \rightarrow 0$.} If a low entropy singularity is an improbable beginning on General Relativistic models, and we know that the universe began in a low entropy state, then some revision to the General Relativistic models is required in which the universe's beginning is not improbable. To produce a model like that, Penrose proposes a mechanism by which preceding physical states could dynamically produce the low entropy condition of the early universe.

As Penrose interprets current fundamental physical theories, length and time scales are determined by the presence of mass in the universe.\footnote{Penrose favorably cites \cite{RughZink:2009} for their relationist view of space and time scales. Also see \cite{RughZink:2017}.} Penrose argues that length and temporal scales ultimately depend upon the existence of mass, so that in a universe in which there are no masses, length and temporal duration lose meaning. If length has lost its meaning, then an infinitely compressed point -- that is, the low entropy initial singularity -- cannot be distinguished from an infinitely large universe.\footnote{I'm speaking loosely. In the absence of length and time scales, a single point and three dimensional space \emph{do} differ, for example, in topological structure. A single point has the topology of $\mathbb{R}^0$ while a three dimensional space has the topology of $\mathbb{R}^3$. But note the qualifications that I made in footnote \ref{singularity_footnote}. For any $a(t) > 0$, three dimensional slices of space-time have the topology of $\mathbb{R}^3$. Penrose should be interpreted as arguing that when length loses its meaning, space-time loses length and time scales. For that reason, we can identify an arbitrarily ``compressed'' three dimensional space with an arbitrarily ``expanded'' three dimensional space.} According to Penrose, mass can be expected to exit our universe for three reasons. First, the universe that we inhabit will expand indefinitely into the future. As our universe expands indefinitely into the future, the density of the universe will decrease and mass will leave our cosmological horizon. Second, some proportion of the mass will be swallowed by black holes and those black holes will decay. Third, Penrose postulates that all of the remaining massive particles will eventually decay into massless products \cite[153]{penrose:2012}. The mass-free homogeneous and isotropic universe towards which our universe tends in the infinitely far future will then be the smooth Big Bang of a subsequent universe. The beginning of a universe would involve a low entropy state because some of the processes that eliminate the masses within a given universe reduce the entropy of the universe.\footnote{\label{nonunitary_footnote}For example, one of the processes Penrose discusses for eliminating masses from the universe involves massive particles being swallowed by black holes and then the black holes undergoing a non-unitary decay process \cite[186-188]{penrose:2012}. The non-unitary decay process reduces entropy. Craig and Sinclair reply to this feature of CCC with the complaint that the late time evolution of the universe will be dominated by the entropy associated with the universe's horizon. This entropy is far larger than the entropy reduced through the decay of black holes \cite[120-121]{CraigSinclair:2012}. While Penrose preempts this objection, Craig and Sinclair (wrongly) complain that Penrose only offers an instrumentalist interpretation of the entropy associated  with the universe's horizon. Importantly, Penrose provides a reply both to the realist and instrumentalist interpretations of the universe's horizon entropy; on Penrose's view, even if the entropy of the universe's horizon is real, that entropy can be ignored because it plays no role in the universe's dynamics \cite[202]{penrose:2012}. Moreover, Craig and Sinclair are inconsistent in their interpretation of the entropy associated with the universe's horizon. As Craig and Sinclair write in their \citeyear[155]{CraigSinclair:2009}, the universe's horizon differs from the horizon of a black hole because the former should not (in their view) be understood as objectively real. But if the universe's horizon is not objectively real, in what sense can the entropy associated with that horizon be understood as objectively real? In any case, the resolution of this debate is irrelevant for my purposes here because I am only concerned with how we ought to interpret CCC.}

In CCC, events in the universe on one side of the interface explain events on the other side of the interface but not vice versa. First, like the anti-inflationary cyclic model discussed in section \ref{AntiInflationSection}, CCC postulates that features of the Cosmic Microwave Background often solved through an inflationary phase in the present universe (e.g., the horizon problem) are instead solved by exponential expansion in a previous universe \cite[210]{penrose:2012}. Second, in addition to reproducing several of the predictions of inflationary  cosmology, Penrose (and collaborators) have argued that the universe prior to ours should leave a signature in the Cosmic Microwave Background not predicted by inflation (\cite[211-219]{penrose:2012}; \citet{An:2018, Gurzadyan:2010, Gurzadyan:2013}).

Photons and gravitons do not possess mass, so the elimination of all \emph{mass} within a given universe would not result in the elimination of all particles. Because photons and gravitons do not have mass, photons and gravitons do not experience temporal duration. Consequently, the trajectory of a photon or a graviton can extend from the present universe to the birth of another universe in the infinitely far future. For example, Penrose argues that collisions between black holes in a previous universe should have left a signal detectable by us \cite[215]{penrose:2012}.

Philosophers of physics distinguish the entropic arrow of time from a variety of other arrows of time -- for example, the temporal asymmetry of causation is referred to as the ``causal arrow of time''. The causal arrow of time has often been understood to align with the entropic arrow of time because both arrows can be afforded a reductive explanation within statistical mechanics (\cite{albert_2000, Albert:2015, Loewer_2007, Loewer:2012, Loewer:2020, papineau_2013}). However, the causal and entropic arrows of time come apart in Penrose's model, provided that what he, and co-authors, claim about the model is correct. That is, the causal arrow of time does not reverse at the interface, even though the entropic arrow of time does reverse.\footnote{Penrose has been careful to argue that while the total entropy of the universe is reduced prior to the bounce through non-unitary processes (see footnote \ref{nonunitary_footnote}), a thermodynamic arrow of time is nonetheless preserved and never reverses direction \cite[175-190]{penrose:2012}. For this reason, Penrose argues that the entropy reduction in CCC is not a violation of the second law of thermodynamics and CCC might not contradict ALP. Be this as it may, if we understand the entropic arrow solely in terms of the entropy gradient, then, according to CCC, there is an epoch in which the entropy gradient points contrary to the causal arrow of time.} Instead, the causal arrow of time is continuous through the interface -- features of a prior universe explain features of a subsequent universe but not vice versa -- and this can be taken to suggest that the direction of time is continuous through the interface.

Craig and Sinclair have provided another reason to think that, in CCC, the interface between universes is the beginning of our universe. As Craig and Sinclair point out, in Penrose's model, some of the mathematical structure usually attributed to time disappears at the interface between universes. For example, in the orthodox interpretation, time scales lose meaning in both the early and late universe. Craig and Sinclair interpret this aspect of the model to mean that the two universes cannot stand in relations of \emph{before} and \emph{after} because to say that one universe preceded the other, when time has lost its meaning at the interface between the two, is incoherent. Instead, Craig and Sinclair claim that there are only topological relations -- and not temporal relations -- between the two universes \cite[127-128]{CraigSinclair:2012}. Here, Craig and Sinclair move too quickly.

In Penrose's model, there are null geodesics that connect the present universe to events in the past universe. Craig and Sinclair agree that the geodesics extend through both universes. But the existence of geodesics traversing the two universes entails that my causal past includes a patch in a previous universe. (For example, photons and gravitons traverse null geodesics between the two universes, e.g., \cite[157-159]{penrose:2012}.) This is precisely the sort of thing that needs to be invoked in order to explain the features that CCC predicts appear in the CMB. The order of events into the causal past, the causal future, and the absolute elsewhere are maintained even when length and time scales are lost because, as Penrose points out, the light cone structure is maintained through the interface \cite[139-147]{penrose:2012}. So, the relations that exist between universes in Penrose's model allow one to make sense of the claim that one of the two universes temporally succeeds the other.

The fact that Craig and Sinclair have misconstrued the consequences of the breakdown of metrical structure is somewhat perplexing, for, elsewhere, Craig has offered a series of objections to a theological view that bares some similarities to Penrose's cosmology and does not make the same mistake in that context. According to the theological view, prior to creating the universe, God existed in metrically amorphous time, that is, that although God's successive mental events (for example) were ordered into relations of before and after, there was no objective fact as to the ratio in the lengths of the non-overlapping temporal intervals occupied by God's distinct mental events (see chapter 9 in \cite{craig_2001:GTE}). Meanwhile, Penrose describes how Weyl maintained a view similar to metrically amorphous time, in which the choice of time scale is a choice of gauge (\cite[451]{penrose:2004}). Because the choice of gauge is conventional and does not correspond to any physical fact, Weyl maintained that there are no facts about time scale. Einstein objected that the conjunction of quantum mechanics and relativity -- by equating energy to both mass and frequency -- suggests time scale is fixed for the rest frame of a given mass. For that reason, Einstein argued that time scale cannot be purely conventional (\cite[453]{penrose:2004}). Hence, Penrose's CCC endorses Einstein's notion that mass fixes the time scale while simultaneously endorsing the view that, without mass to fix the time scale, time would be metrically amorphous. Though Craig argues against the view that God existed in a metrically amorphous state prior to creation, Craig admits that, unlike a timeless state, a metrically amorphous state \emph{can} stand in the before relation with respect to the universe: ``this [metrically amorphous] state exists literally before God’s creation of the world and the inception of metric time'' (\cite[270]{craig_2001:GTE}). So, just as a deity who is metrically amorphous prior to creation can stand in the before relation with respect to that deity's creation, so, too, can a metrically amorphous physical state stand in the before (or after) relation with respect to metric time.

This point is worth unpacking in some more technical detail. Readers who are satisfied with the qualitative description already given can safely skip to the next section. Any metric tensor $g_{\mu \nu}$ can be decomposed into a volume element (i.e, $|\det(g_{\mu\nu})|^{1/4}$) and a conformal metric density (i.e., $\tilde{g}_{\mu \nu}$) (as first worked out in detail in \cite{Yerkes:1925, Yerkes:1932A, Yerkes:1932B}; also see \cite[63-64]{Anderson:1967}; \cite{Anderson:1971}). That is, $g_{\mu \nu} = \tilde{g}_{\mu \nu} |\det(g_{\mu\nu})|^{1/4}$. Penrose maintains the \emph{Weyl Curvature Hypothesis} (WCH), according to which the Weyl curvature tensor exactly vanishes in the low entropy condition of the early universe (\cite[132-135]{penrose:2012}). If the Weyl curvature vanishes everywhere on a region of space-time, then that region is conformally flat, that is, there exists a conformal transformation to a diagonal (flat) metric (\cite[63]{Anderson:1967}; \cite[464]{penrose:2004}). This follows as a consequence of the fact that the Weyl curvature tensor can be expressed entirely in terms of the conformal metric density and its inverse and is equal to the Riemann curvature tensor formed by substituting $g_{\mu\nu}$ with $\tilde{g}_{\mu \nu}$ \cite[63-64]{Anderson:1967}. For example, the Weyl curvature tensor vanishes in FLRW space-times and, as a result, there exists a conformal transformation from FLRW space-time to Minkowski space-time. Thus, if the WCH is true, then there exists a conformal transformation from the early universe to a flat space-time (\cite[464]{penrose:2004}). According to CCC, in the far future, when all mass in the universe vanishes, the universe will again be conformally flat. Thus, the early universe, the late universe, and a corresponding region of flat space-time are all conformally equivalent; that there is no mass present at early or late times entails that all three are physically equivalent. So, Penrose concludes, conformal structure remains at early and late times even though the full metrical structure is lost.

At early times, the volume element vanishes -- reflecting the singularity in classical models of the Big Bang -- but the conformal metric density is perfectly well-behaved. Since CCC attributes physical significance only to those features invariant under conformal transformations in the early or late universe, the volume element -- and so the associated singularity -- has no physical significance.\footnote{Thanks to an anonymous referee for pointing this out.} In turn, the conformal metric density can be used to smoothly continue time-like curves from one universe into the next. As is well known, conformal transformations are precisely those that leave the light cone structure invariant; importantly, this has the implication that, under the conformal transformation, future (past) directed tangent vectors are mapped to future (past) directed tangent vectors. Thus, the light cones along any given time-like curve encodes a conformally invariant temporal ordering of events. Because the conformal metric density allows one to continue time-like curves through the interface between universes, the two universes can be placed into before and after relations, as expected.

\section{Conclusion}

One can understand the relationship between cosmological models and the \emph{Kal$\overline{\textrm{a}}$m} argument in one of two ways. First, one might think that the \emph{Kal$\overline{\textrm{a}}$m} argument is strongly supported by a priori arguments, alone. In that case, the \emph{Kal$\overline{\textrm{a}}$m} argument might be defended by merely demonstrating the logical consistency between the beginning of the universe and our best cosmological models. However, many regard the a priori case for the \emph{Kal$\overline{\textrm{a}}$m} argument as particularly weak. Second, one might think that the \emph{Kal$\overline{\textrm{a}}$m} argument is jointly supported by a priori and empirical arguments. In that case, one cannot merely show that there exists an interpretation of a cosmological model that is logically consistent with the universe having begun to exist. Instead, one must show that the interpretation is ultima facie plausible. Consequently, if Craig and Sinclair intend for their argument to be jointly supported by a priori and empirical argumentation, then Craig and Sinclair need to meet a much higher burden of proof than Craig and Sinclair have typically met thus far.

We've seen that Craig and Sinclair's interpretation of bounce cosmologies does not sit well with a number of features of bounce cosmologies. On the one hand, there are cyclic models, like those endorsed by Ijjas, Steinhardt, Turok, and Penrose, or models in which universes are born out of black holes, like Smolin's or Poplawski's, in which features of one universe explain features of a subsequent universe. We've also seen that there are a number of bounce cosmologies in which the entropy is ``reset'' at the interface even though the thermodynamic arrow of time is continuous along time-like and null geodesics piercing the interface. Whether any of the cosmological models I've discussed here are plausible continues to be discussed by physicists. Nonetheless, provided that one of these models -- or a model appropriately similar -- does turn out to be strongly confirmed by the evidence, then Craig and Sinclair's contention that the universe began to exist will have lost considerable empirical support because, as I have shown, bounce cosmologies should not be interpreted as depicting the metaverse as having begun to exist.

\bibliographystyle{plainnat}
\bibliography{references.bib}

\end{document}